# Recommender System for News Articles using Supervised Learning

Akshay Kumar Chaturvedi

(MIIS Master Thesis)

**Supervised by:**

Dra. Filipa Peleja

(Vodafone Portugal)

Dra. Ana Freire

(Department of Information and Communications Technologies - Universitat Pompeu Fabra)

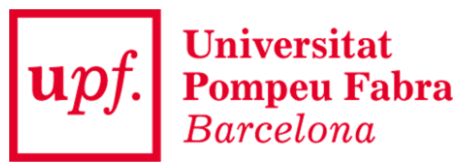

# Abstract

In the last decade we have observed a mass increase of information, in particular information that is shared through smartphones. Consequently, the amount of information that is available does not allow the average user to be aware of all his options. In this context, recommender systems use a number of techniques to help a user find the desired product. Hence, nowadays recommender systems play an important role.

Recommender Systems' aim to identify products that best fits user preferences. These techniques are advantageous to both users and vendors, as it enables the user to rapidly find what he needs and the vendors to promote their products and sales. As the industry became aware of the gains that could be accomplished by using these algorithms, also a very interesting problem for many researchers, recommender systems became a very active area since the mid 90's.

Having in mind that this is an ongoing problem the present thesis intends to observe the value of using a recommender algorithm to find users likes by observing her domain preferences. In a balanced probabilistic method, this thesis will show how news topics can be used to recommend news articles.

In this thesis, we used different machine learning methods to determine the user ratings for an article. To tackle this problem, supervised learning methods such as linear regression, Naive Bayes and logistic regression are used. All the aforementioned models have a different nature which has an impact on the solution of the given problem.

Furthermore, number of experiments are presented and discussed to identify the feature set that fits best to the problem.

# List of Figures

# List of Tables

# Table of Contents

# 1. Introduction

This is the age of information and consumption of information has gained pace, particularly in recent years. People in today's world are overwhelmed with data and this often creates the problem of choosing something from a large set of options. Recommender systems identify which products should be presented to the user, in which the user will have time to analyse and select the desired product [Ricci et al., 2011].

Now with the advent of e-commerce websites like Amazon, it became more obvious the important role that recommender systems play. The importance of recommender systems also came to light in 2006. In 2006, Netflix, a global provider of streaming movies and TV series, announced an open competition to predict user ratings for films, based on previous ratings without any information about the users or the films. Prizes were based on improving Netflix's own algorithm. The team that finally won the competition was able to achieve around 10% improvement on Netflix's algorithm [Ricci et al., 2011], they were awarded 1 million US$ .

A recommender system is traditionally composed of:

- **Users** : People in the system who have preferences for items and people who can be source of data as well are called Users [Ricci et al., 2011]. Each user may have a set of user attributes, if we are using user demographic (age, gender etc.) then demographic is a user attribute. For each user, a model can be inferred. For example, their genre of movie or the type of books they (the users) like to read.

- **Items** : Products the system is choosing to recommend are known as Items [Ricci et al., 2011]. For each item we may have a set of item attributes or properties. For example, an actor in a movie, author of a news article or colour of an appliance.

- **Preferences** : These represent users' likes and dislikes [Ricci et al., 2011]. User meets item in the preferences space, for example, a user can rate a movie 5 on a scale of 5 stars.

The components described above are used in different algorithms in various ways. Broadly these algorithms can be classified in the following types:

- **Non-personalized systems** : These involve summary statistics and in some cases product associations [Poriya et al., 2014] by using external data from the community, like, a product that is the best seller or most popular or trending hot. It may also provide summary of community ratings, for example, how much a population likes a restaurant or summary of community ratings which turns into a list like, which is the best hotel in town.

- **Content based filtering** : Here, users rate items and from that a model of user preferences against the item attributes is built [Ricci et al., 2011]. An example could be in the domain of movies. Suppose someone likes science fiction, fantasy and action movies, and doesn't like romantic movies. Overtime the algorithm can accumulate this and Figure out that the user has positive scores on genres like science fiction, fantasy and action, and lower scores for romance. The algorithm might also find out that there were some actors that user likes or dislikes. For example, the user can be a fan of movies with the actor Bruce Willis, and not a fan of Ben Stiller. Content based filtering uses this information to map user ratings against the attributes of the products - in this example movies. Our approach fits in this type of algorithm as we are trying to build a model on user feedback of news articles. Figure 1.1 provides a good overview of Content Based Filtering.

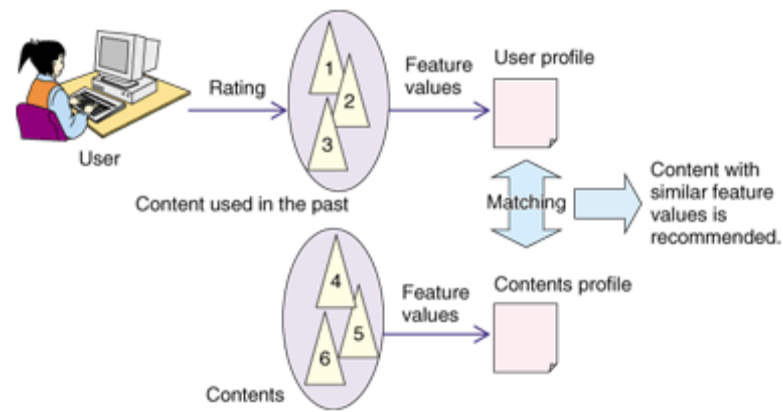

Figure 1.1 : Content Based Filtering

- **Collaborative filtering** : In collaborative filtering, user ratings of other people are used rather than attribute data to predict and recommend [Ricci et al., 2011]. Collaborative filtering builds on the idea of a user model that is a set of ratings and an item model that is a set of ratings. Combining the two models, we get a sparse matrix of ratings, some of its cells are filled and most are not. So, here there are two main tasks, one is to fill the empty cells or predict a rating and second is to choose a filled cell or recommend an item. A good overview of Collaborative Filtering is presented in Figure 1.2.

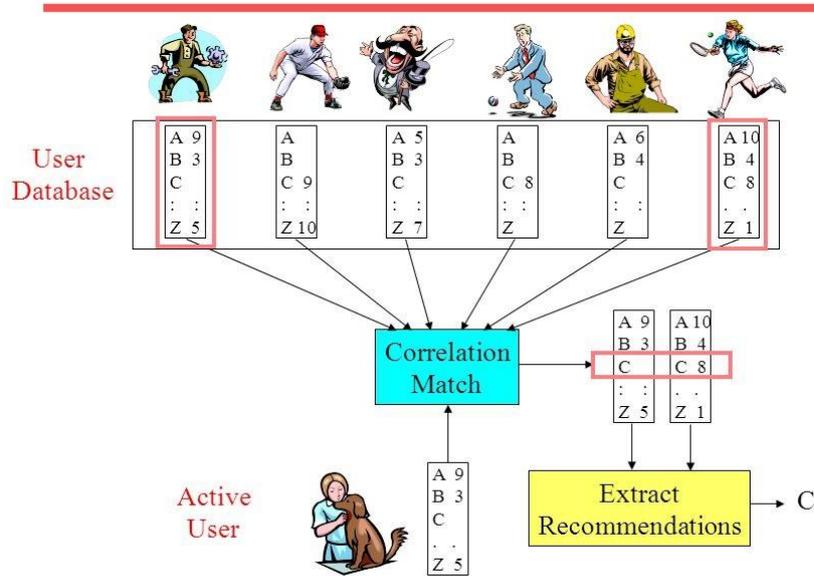

Figure 1.2 : Collaborative filtering

The main difference between Content Based Filtering and Collaborative Filtering is that Collaborative Filtering works on preferences of other users (users with similar preferences for some items) to recommend new items whereas Content Based Filtering is not at all concerned with preferences of the other users. This point is appropriately illustrated in Figure 1.3.

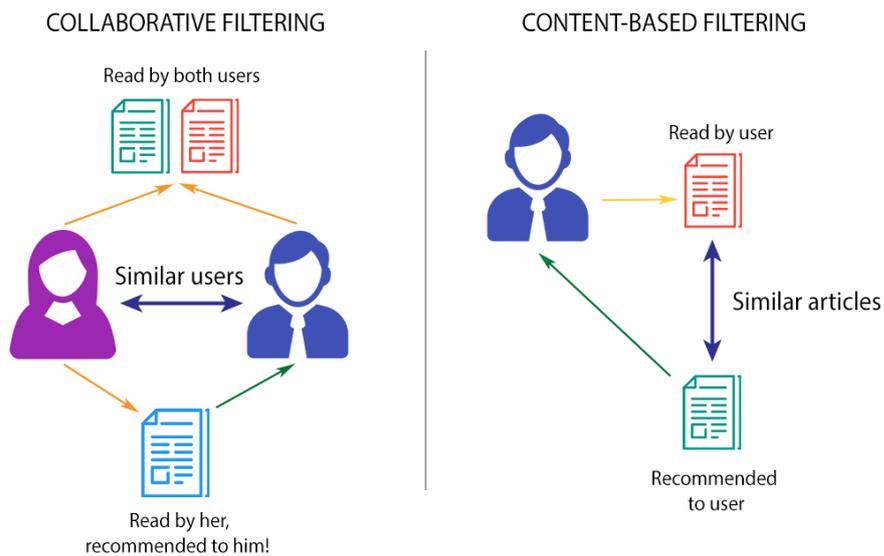

Figure 1.3 : Collaborative Vs Content Based Filtering

- **Hybrid methods** : In these methods, a combination of two or more recommendation algorithms are used to take or maximize advantage of some techniques and avoid or minimize the drawbacks of another [Burke, 2002]. Various ways of combining different algorithms are shown in Table 1.1 (for more details please see [Burke, 2002]).

Table 1.1 : Methods of Hybridization

| Hybridization method | Description |
| --- | --- |
| Weighted | The scores (or votes) of several recommendation techniques are combined together to produce a single recommendation. |
| Switching | The system switches between recommendation techniques depending on the current situation. |
| Mixed | Recommendations from several different recommenders are presented at the same time |
| Feature combination | Features from different recommendation data sources are thrown together into a single recommendation algorithm. |
| Cascade | One recommender refines the recommendations given by another. |
| Feature augmentation | Output from one technique is used as an input feature to another. |
| Meta-level | The model learned by one recommender is used as input to another. |

# 2. Problem definition and thesis objective

People tend to have preferred sources for printed news, like a newspaper or a magazine (Figure 2.1 shows the news recommendation section of The New York Times). However, in online world, this reality changes drastically, a user is flooded with information from different sources in which it is not uncommon for a user to change between news portals or read news from portals that merge news articles from different sources (Yahoo! News). Having such a huge amount of information, it becomes difficult to select the news article that a user will like. Consequently, users stop the consumption of news or lowers it down.

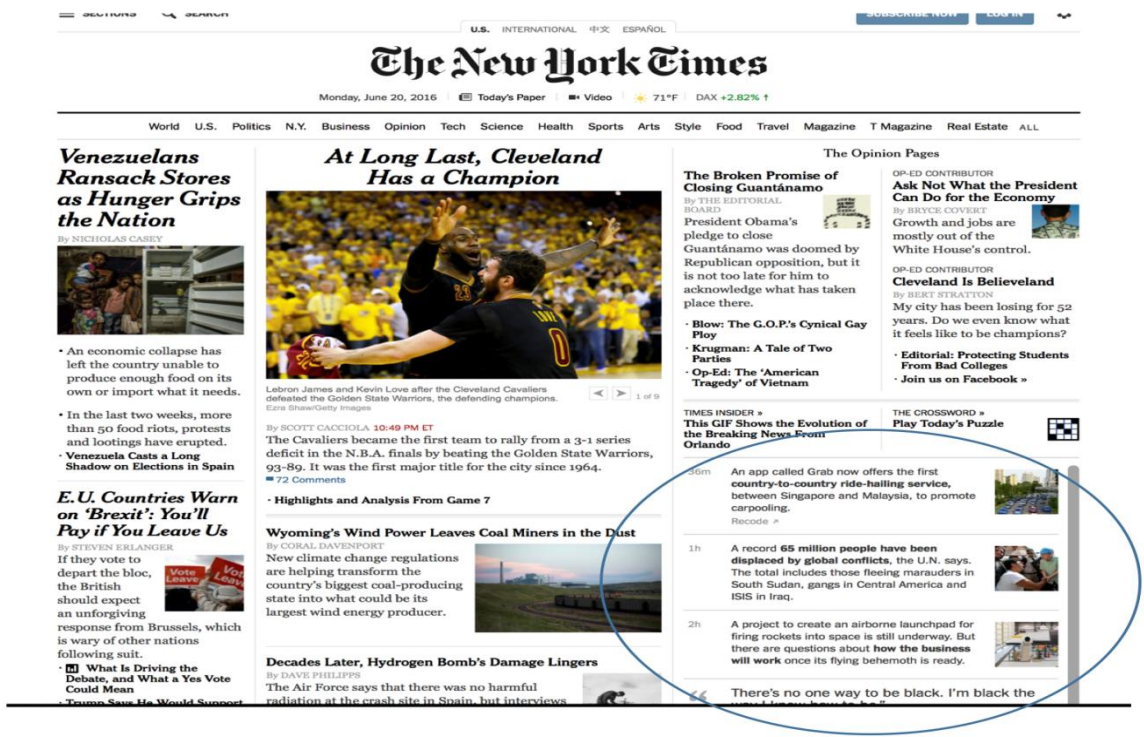

Figure 2.1 : The New York Times' recommendation section.

To overcome the problem of information overloading (difficulty in making a decision caused by the presence of too much information), recommender systems can be used. These types of systems use various types of feedbacks from the users to narrow down the users' options to a small number. Rating an article is the most common feedback that is used, but ratings alone are not always sufficient to predict the articles a user might like. So, other types of feedbacks are used to improve the recommendation. One can try various implicit feedbacks as user is not asked to provide them. But when it comes to explicit feedback, there are a few things one has to take into account, firstly, the total number of explicit feedbacks used by the recommender system should be minimum, as users usually don't have much time or don't like to provide feedbacks. Secondly, feedback asked to the user should be easy for an average user to understand so that he can provide the feedback easily and correctly.

The goal of the present thesis is to investigate the impact of features extracted (explained in section 6.1) from feedback 'classes' (category a news article belongs to, for example, sports, politics, etc.) on recommendations and also on other features. To achieve this goal, we used the extracted features in machine learning models with different feedbacks in the dataset. We have further created new datasets by selecting the articles belonging to the most frequent classes and removing the articles belonging to the infrequent ones' and then performed experiments on them to see the impact of number of classes on the performance of the system. At last, we have done analysis of the results obtained from using feature combinations in different models to figure out the right model and features required to make reasonable recommendations.

# 3. Problem Formalization

In this section, we will formalize our problem and also give a brief overview how we achieved our objective. Our problem of recommending news articles can be formalized as follows: given feature(s) F, learn a model which can estimate user ratings R, such that mean squared error (MSE) between original (R) and estimated value (r) is minimal, where

- F represents the features, it can be visualized to be a matrix of size M by N, where N is the number of different types of feedbacks taken from user and M is the number of articles. It contains every type of feedback taken from user except user ratings (R), which is to be estimated. We can use all the features or select a subset.

- R is the user ratings given by user for each article. Domain of user ratings can be binary or from a range of integers 1 to 5 or 1 to 10, in our dataset it is 1 to 5.

- r is the estimated user rating (we find 'r' using supervised learning algorithms in the present thesis) for each article. It takes the same values as R.

- MSE or mean squared error is defined as $\frac{1}{M}\sum (R-r)^2$, M is the number of articles.

To achieve the above objective we experimented with three Supervised Learning algorithms, namely Linear Regression, Logistic Regression and Naive Bayes. The three algorithms are described in sections 4.1, 4.2 and 4.3. Experiments were conducted not only with the features available in the data set, but also with two extracted features which are described in section 6.1. After obtaining results (MSE) from our three algorithms mentioned before, these results are compared with the result of a baseline model which is also described in section 6.1.

# 4. Background

Machine Learning can be described as inferring or learning models from data itself about the process that generated the data. A more formal definition was given by Tom M. Mitchell in [Mitchell, 1997] : "A computer program is said to learn from experience E with respect to some class of tasks T and performance measure P if its performance at tasks in T, as measured by P, improves with experience E". Machine Learning can be broadly classified into three types:

**Supervised Learning** : When the given data contains the output explicitly and the pairs of input and output data points are used to predict the unknown output of inputs, it is called supervised learning [James et al., 2014] . An example could be a movie data set with different features of movie like director, lead actor or genre (as inputs) and also user rating (as output).

**Unsupervised Learning** : Instead of having input with correct output, data only contains input. Here, the objective is to cluster or group the similar inputs together [James et al., 2014] . An example could be clustering documents in different genres given the most frequent words in each document.

**Reinforcement Learning** : Here, situation is a bit similar to unsupervised learning in the sense that it does not use input with correct output, instead it uses a reward for every action [Mitchell, 1997] . The rewards can be both positive and negative depending on the action taken given the input. So, here the objective is to maximize the total reward which is the sum of all rewards which the system got after every action. An example could be playing a game where for every right move a positive reward is given and for every wrong move a negative reward is given, thereby the objective is to figure out the set of moves or actions which maximizes the total reward.

Most of the machine learning algorithms fall into one of the three categories described above. We will describe below some of the algorithms but will concentrate only on supervised learning methods as it is the relevant type in this thesis.

## 4.1. Linear Regression

Linear Regression has two main objectives : the first one is to establish if there is a relationship between two variables. This relationship can be positive i.e. both variables increase and decrease together, or negative i.e. when one variable increases, the other decreases and vice-versa. The second objective is to forecast output values for new inputs. The output of this algorithm is a linear equation,

$$y = \beta_0 + \beta_1 * x + \acute{\varepsilon} \qquad (1)$$

where y is the dependent variable or the output the model will predict and x is the independent variable or the input. $β_0$ is the constant or intercept of the equation, it doesn't have an intuitive interpretation. $β_1$ is the slope and is the relationship between variables, it is the amount by which y changes when x changes. In most cases, data is not completely linear i.e. some data points may fit the line exactly and some may not, in that case, we get error between actual output and predicted output, $έ$ is that error (This error is shown in Figure 4.1). Gradient descent algorithm is used to find optimal values of $β_0$ and $β_1$ which minimizes error $έ$, thereby finding the best line to fit the data [James et al., 2014] . In case, data is not at all linear then linear regression won't work.

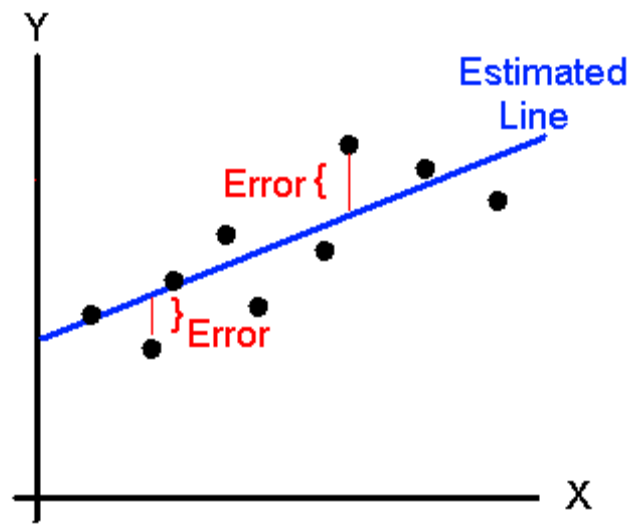

Figure 4.1 : Linear Regression

Linear Regression can be of two types, simple linear regression and multiple linear regression. In simple linear regression, there is only one independent variable (x). In multiple linear regression, there are more than one or multiple independent variables ($x_1, x_2, x_3....$). In multiple linear regression, a subset of variables can be selected rather than using all of them. Two strategies are used for this selection, one is to check the relationship of different independent variables with dependent variable (y) and select only those independent variables which are strongly related to y. Second strategy is based on overall model performance, here the contribution or impact of an independent variable is checked i.e. whether it is able to help in predicting outputs or not.

## 4.2. Logistic Regression

Logistic regression is also a linear model. It differs from linear regression in the context that in linear regression, the dependent variable is continuous whereas in logistic regression, the dependent variable is discrete. Most of the times the dependent variable is binary discrete, having value 0 or 1 i.e. falling into one category or another. The output of logistic regression is probability of a input belonging to one category or another [James et al., 2014] . Linear equation used in linear regression is used here also, but here linear equation is passed into a logistic function (logistic function is shown in figure 4.2).

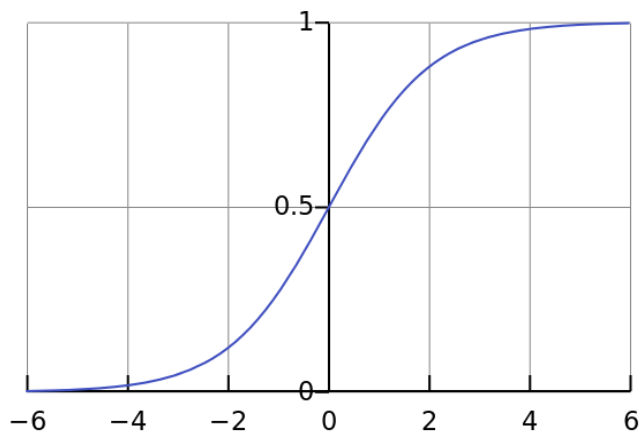

Figure 4.2 : Logistic Function

Thereby, giving the following equation :

$$P(y = 1 \mid x) = \frac{1}{1+e^{-(\beta_0 + \beta_1 * x)}} \qquad (2)$$

where y is the output, x is the input and $\beta_0 + \beta_1*x$ is the linear equation used in linear regression. A threshold is selected, if the probability is above the threshold, input belongs to one category, if the probability is below the threshold, it belongs to another. Threshold is adjusted according to number of points classified correctly, most common threshold is 0.5. A question could be, why use logistic function? The answer is that here the output is probability is probability and probability is always between 0 and 1, to restrict the output between 0 and 1, logistic function is used.

## 4.3. Naive Bayes

Naive Bayes like logistic regression predicts if an input belongs to a particular category or not. It's output is also probability but instead of logistic function, it uses Bayes rule. Figure 4.3 describes Bayes rule appropriately :

$$P(c|x) = \frac{P(x|c)P(c)}{P(x)}$$

where $P(c|x)$ is the Posterior Probability, $P(x|c)$ is the Likelihood, $P(c)$ is the Class Prior Probability, and $P(x)$ is the Predictor Prior Probability.

$$P(c|X) = P(x_1|c) \times P(x_2|c) \times \cdots \times P(x_n|c) \times P(c)$$

Figure 4.3 : Bayes Rule

So the equation is :

$$P(y = 1 | x) = \frac{P(x|y=1)*P(y=1)}{P(x)} \qquad (3)$$

where y is output and x is input. P(x | y) and P(y) are calculated from the given data and used for predicting outputs of new inputs. For different values of y, P(y | x) is calculated and x is assigned to the category (or y is assigned the value) for which P(y | x) is maximum [Rish, 2001] . An assumption is made that variables $x_1, x_2$.... are independent of each other [Rish, 2001] and therefore, $P(x_1, x_2.... | y)$ can be written as $P(x_1 | y) * P(x_2 | y)$....., which is not true is most cases and that's why this algorithm is called Naive Bayes.

# 5. Related work

Recommender systems have been of great interest among the scientific community in the last few years. Several successful approaches have been proposed. These approaches can be classified by the domains they have been used in or applied to. Approaches are further classified by the type of data used for recommendation (implicit or explicit) and lastly by the type of algorithm used for recommendation (collaborative, content based filtering or hybrid approach). In this section we will go through some papers which have proposed recommender systems in domains like news, music etc. using both explicit and implicit feedback or either of them in different algorithms.

Amongst the many papers which both inspired and motivated me to work on this topic, one was [Li et al., 2010]. Here, news recommendation problem was modelled as a contextual bandit problem. Authors devised a new algorithm named as LinUCB to solve the problem. LinUCB is a general contextual bandit algorithm and can be applied to domains other than news recommendation. Using LinUCB, number of clicks increased by 12.5% as compared to a standard context-free bandit algorithm.

While talking about news recommendation, one cannot miss [Das et al., 2007]. Here, both model and memory based algorithms have been used. Memory based algorithms use weighted average of past ratings from other users where weight is proportional to the 'similarity' between users. Pearson correlation coefficient and cosine similarity are the typical measures used for 'similarity'. Model based ratings model the user preferences based on past information. From model based, they used clustering techniques PLSI and MinHash, and from memory based, they used item co-visitation. All the three algorithms assign a score to a story. Finally, two combinations of three techniques (PLSI, MinHash and co-visitation) were used for evaluation. In one combination, co-visitation was given a higher weight than the other techniques (2.0 instead of 1.0). This method was named CVBiased and in another combination PLSI and MinHash were given higher weight (2.0 instead of 1.0), which is known as CSBiased. The baseline used was recommendation based on recent popularity called Popular. It was observed, on average both CVBiased and CSBiased performed 38% better than the baseline Popular on live traffic.

Another influential paper is [Liu et al., 2010] , the work done in this paper was built on [Das et al., 2007] . Here, information filtering was used. Information filtering removes unwanted information from an information stream. First, they did an analysis of a users' interest over a 14 month period for each category of article (categories are pre-defined). They used click distribution of a user to study the interest. Then they used Bayes rule to predict a users' interest for a particular period of time. Then, predictions made for particular period were combined to predict a users interest in a long period. The predictions made till now were users' genuine interest. But a user also gets influenced by current news trends, so Bayes rule is used to make predictions using click distribution of a short recent period (example last hour). Finally, users' genuine interest is combined with current news trends to get IF (article).

The information filtering score for each article IF (article) is multiplied by collaborative filtering score for the article CF (article), and the final score after multiplication of IF (article) and CF (article) is used to rank the articles. The collaborative filtering score CF (article) is obtained from [Das et al., 2007] . Using this method, click through rates (CTR) were improved by 30.9% upon existing method i.e. [Das et al., 2007] .

While papers discussed previously were concerned with news domain in particular, [Hu et al., 2008] is not concerned with any particular domain, instead here, focus was on recommendations using implicit feedback dataset. Authors used latent factor models to compute the number of times a user 'u' watched a program 'i' ($r_{ui}$). For finding the parameters to compute $r_{ui}$ , alternating least squares optimization technique was used with regularization to avoid overfitting. This approach was to recommend television shows. The interesting thing about this approach was there was only positive feedback and no negative feedback and this made the task of recommendation much more difficult. Evaluation was done by the users themselves as they were provided a list of television shows sorted from most preferred to least preferred shows. Expected percentile rank was used for comparison where latent factor model outperformed both popularity model and neighbourhood model.

In [Baltrunas and Amatriain, 2009], micro-profiling of users was used for music recommendation. Several sub profiles of each users were created. Instead of a single user model several models were created to make recommendations with different context. In this paper, context was time. So several sub-profiles were created for morning, evening, summer, winter, etc. The idea behind these sub-profiles was that music preference of a user varies with time but it also remains same for a particular time. To make it more clear, an example could be, one might like inspirational music in the morning and romantic music in the evening so there is a difference between morning and evening music preference but music preference for morning remains same, it doesn't change from inspirational to any other type. Only implicit feedback was used here, so as in [Hu et al., 2008], dataset only had positive opinions and not negative. For training, popularity of an artist was considered, i.e. number of times a user has listened to a particular artist at a particular time domain (morning, evening, etc.). In this paper, artists were recommended rather than a music track. Offline evaluation was done, where mean absolute error was computed for each time domain. And then, a weighted average of mean absolute error was taken where weight was the number of ratings that can be predicted in a particular time domain. The best system would be the one which could minimize the weighted average of mean absolute error. In conclusion, it was observed that contextual (different time domains) micro-profiling gave better results than a context free user model.

In [Parra and Amatriain, 2011], focus was again on mapping implicit feedback to explicit ratings. The domain was music recommendation. The contributions of this paper are, a study of the relation between implicit feedback and explicit ratings, analysis of impact of other variables and a linear model to predict ratings. A user study was conducted to collect data, users were asked to rate an album on a 1 to 5 scale. Three variables were selected for analysis, implicit feedback (play count for a user on a given item), global popularity (aggregated play count of all users on a given item) and recentness (time elapsed since user played a given item). Values of all the three variables were divided into three bins - low, medium and high. It was observed that quantized implicit feedback and distribution of ratings were related, quantized recentness and distribution of ratings were also related but not as strongly as implicit feedback and no significant relation was found between quantized global popularity and distribution of ratings. Analysis of relationship between other variables showed some relation between recentness and implicit feedback only and not any other pair. Linear regression was used for predicting ratings using three variables and RMSE (root mean squared error) was used for evaluation. Four models were used for prediction, first one was only with implicit feedback, in second one recentness was also added with implicit feedback, in third model global popularity was added to the second model and in the fourth and last model a variable which was multiplication of implicit feedback and recentness was added to the third model. the baseline used was average of ratings. Although performance of all models were relatively same, but they all had average improvement of 6.5% over the baseline.

[Parra et al., 2011] was built upon [Parra and Amatriain, 2011]. In [Parra et al., 2011], instead of linear regression, logistic regression was used for prediction. Here, a random effect of user was also added in the model. Finally, rating is predicted by computing expected value of rating across all ratings and their probabilities for a user and a item. A second dataset was collected which contained only implicit feedback i.e. play count of each album per user. Experiments were performed on this dataset and also the dataset collected in [Parra and Amatriain, 2011] . While conducting experiments on first dataset [Parra and Amatriain, 2011] , a variable, concerts per year and random effect of user was used with the three variables used in [Parra and Amatriain, 2011] . In case of second dataset, as it did not have concerts per year variable in it all the other variables used in first dataset were used. In the end, results of four approaches were compared. First approach was the one used in [Hu et al., 2008], in second approach log transformation of play counts used in [Hu et al., 2008] were used, third approach is the one where logistic regression was used with implicit feedback, recentness and global popularity, in fourth approach concerts per year is also used with the variables used in third approach, fourth approach was only used on first dataset as concerts per year was available for that dataset only. For evaluation, Mean average precision (for relevant albums) and normalized Discounted cumulative gain (for ranking of albums) metrics were used as final goal in this paper was recommending albums rather than just predicting ratings. Most popular album was used as baseline and it outperformed all the other approaches in the first dataset as the dataset was very sparse. In second dataset, results were different, baseline performed worst, second approach improved upon first one and although third approach performed better than first one it was more or less equal to second approach.

Another paper which is built upon [Hu et al., 2008] is [Johnson, 2014]. Here also like [Hu et al., 2008], latent factor models are used, though here the approach is probabilistic which was not the case in [Hu et al., 2008]. Here, the algorithm used was called logistic matrix factorization. The probability of interaction between a user and a item is given by a logistic function parameterized by the sum of the inner product of user and item latent factor vectors and user and item biases. Similar to [Hu et al., 2008], a confidence function is chosen, this function can be replaced any other function, in the paper it is given by $\alpha r_{ui}$ where α is the tuning parameter and  $r_{ui}$ is rating of user u for item i. Then, likelihood of the observation matrix is computed, regularization is also performed using zero mean spherical Gaussian priors to avoid overfitting. Local maximum of likelihood function is found using alternate gradient descent method. As in [Hu et al., 2008], here also play counts are used for prediction, a play count is only recorded if a user has listened to a song for more than 30 seconds continuously thereby removing the bias that a user started a song but didn't like it. Mean Percentage Ranking (MPR) metric was used for evaluation as was the case in [Hu et al., 2008]. Logistic Matrix Factorization was able to outperform both the model used in [Hu et al., 2008] and also baseline popularity model.

For recommending news, [Li et al., 2010], [Das et al., 2007]  and [Liu et al., 2010]  used history of logged-in users, [Garcin et al., 2012] considers only current visit data where users don't log in. Authors decided to build a candidate set of most fresh news articles. Potential candidates for recommendation were the articles in candidate set which were accessed at least once. Most popular news article was taken as baseline. Three approaches were considered, collaborative filtering at the level of news items, content based recommendation and a hybrid approach where collaborative filtering is applied at the level of topics. In the first approach, news reading was modelled as a Markov process given the sequential nature of news reading, news articles were considered states. Two parameters were used at this stage namely past and future, past refers to the number of news articles to be considered for recommendation whereas future refers to the number of steps needed to reach a given state. Transition probability is computed for pairs, using these transition probabilities a score is generated which is reading probability of a news item n in f steps given the sequence s, here the order of sequence matters. In second approach, Latent Dirichlet Allocation (LDA)  was used with cosine similarity. In third and final approach, news articles were clustered based on topic distributions, then a transition matrix is build for clusters. K-means is used for clustering. Once the clusters are determined, a probabilistic model is build similar to first approach by replacing a sequence with a cluster. Two metrics were used for evaluation, Mean average precision (MAP) and S@5, it is equal to 1 if the immediate successor of the current items is recommended among the first 5 recommended news stories, 0 otherwise. The training set contains all news items accessed before time t and the testing set has items accessed after time t. First approach i.e. collaborative filtering at the level of news items outperformed others in both metrics.

There are similarities and differences between the works described above and the work we have done. The similarity and difference is in the evaluation metric, machine learning model and features used in the model. There are papers which have used Naive Bayes [Liu et al., 2010] , linear regression [Parra and Amatriain, 2011]  and logistic regression [Parra et al., 2011]  , but none of them have shown a comparison of these three models. In evaluation, RMSE has been used [Parra and Amatriain, 2011]  but most of them have preferred MAP and other metrics, some have even tested their systems on live traffic. In features, none of them have used joint and mean conditional probabilities of labels givens by users for a news article, however, we have experimented with feature novelty which is similar to feature recentness used in [Parra and Amatriain, 2011]  and [Parra et al., 2011] .

# 6. Experiments

## 6.1. Overview

In this section, we will give an overview of our experiments and also explain the extracted features and baseline used in our experiments. The first step in performing experiments was to collect data, instead of collecting data ourselves (as process of data collection can be both expensive and time consuming), we downloaded data used in a CMU (Carnegie Mellon University) project which seemed suitable for our experiments. The details of the data set are given in section 6.2. Next step was data pre-processing where we removed ambiguities and handled missing values, detail are in section 6.3. After data pre-processing, process of feature extraction was performed, details of how feature extraction was done is in section 6.4. After feature extraction, different features were used in different algorithms (as described in sections 4.1, 4.2 and 4.3) to predict user ratings. In section 6.5, we have presented the results of our experiments.

Although the features available in the data set can be used for news recommendation, we decided to extract features and use them also for news recommendation. Two features were extracted namely Joint Conditional Probability (JCP) and Mean Conditional Probability. The difference in these two features is at the last step of computation. We calculate the conditional probability of individual labels or tokens (given by an user to an article) given (or conditioned upon) user rating of the article. If an article has been given only one token or label, then there will be no difference between JCP and MCP. If an article has been given multiple tokens or labels, then there can be difference between JCP and MCP. The difference is as follows :

**Joint Conditional Probability (JCP)** : Conditional probability of individual tokens or labels are multiplied together and the result of the multiplication of the conditional probabilities of individual tokens or labels is JCP.

**Mean Conditional Probability (MCP)** : For MCP, mean of conditional probabilities of tokens or labels is computed and the mean is known as MCP.

The motivation behind extracting and using these particular features came from [Liu et al., 2010] . Using similar features in Naive Bayes, authors of [Liu et al., 2010]  were able to develop a very efficient news recommender system. Although, [Liu et al., 2010]  was built upon [Das et al., 2007] , what caught our eye was the simplicity of features and methods used in [Liu et al., 2010] . While doing literature review, we couldn't find any paper where these features were used, what we found was people using linear and logistic regression with features like recentness, etc. So we decided to use JCP and MCP in linear regression, logistic regression and Naive Bayes for performing news recommendation.

**Baseline** : For baseline, we used popularity model. In popularity model, the most popular or most frequent value is always predicted. For example, if user rating '5' is most frequent when compared to other values of user rating in the given data set, then each predicted user rating will be 5.

## 6.2. Dataset

The dataset we used for our experiments was collected in Carnegie Mellon University for the project PIIR (Proactive Personalized Information Integration and Retrieval) [Özgöbek et al., 2014]. Original unclean dataset contained information on 10010 user ratings on news articles. Dataset contains both explicit and implicit feedback. Table 6.1 shows the description of explicit feedbacks in the dataset.

Table 6.1 : Description of explicit feedbacks.

| *Feedback* | *Description* |
|---|---|
| *user_like* | An integer between 1 and 5 representing how much a user likes an article. |
| *Relevant* | An integer between 1 and 5 representing how much a user thinks an article is relevant. |
| *Novelty* | An integer between 1 and 5 representing how new the article is according to the user. |
| *Readability* | Representing whether the article is readable or not is according to the user, is either 0 or 1. |
| *Authority* | Representing whether the news article is authoritative or not according to the user, is either 0 or 1. |
| *Classes* | A string representing the classes news article belongs to (for example, |news), an article can belong to multiple classes (|news |movies |Broadway |music), a user can create any class, there are no predefined classes. '|' serves as delimiter. There were no classes for some articles which means the user didn't give a feedback. |

The variables described in Table 6.1, except the variable "classes", contain value -1 for few articles. This means user didn't provide any feedback for that variable in a specific article. Table 6.2 shows the description of implicit feedbacks in the dataset. Table 6.3 and 6.4 gives statistics of both explicit and implicit feedbacks in the data set.

Table 6.2 : Description of Implicit feedbacks

| Feedback | Description |
|---|---|
| *TimeOnMouse* | Number of milliseconds the user spent on moving the mouse. |
| *TimeOnPage* | Number of milliseconds the user spent on a page or news article. |
| *EventOnScroll* | Number of clicks on the scroll bars. |
| *ClickOnWindow* | Number of clicks inside browser window but not on scroll bars. |
| *TimeOnHScroll* | Number of milliseconds the user spent on using horizontal scroll. |
| *TimeOnVScroll* | Number of milliseconds the user spent on using vertical scroll. |
| *NumOfPageUp* | Number of pages the user scrolled up. |
| *NumOfPageDown* | Number of pages the user scrolled down. |
| *MSecForPageUp* | Number of milliseconds the user spent on scrolling a page up. |
| *MSecForPageDown* | Number of milliseconds the user spent scrolling a page down. |
| *NumOfUpArrow* | Number of clicks on up arrow key. |
| *NumOfDownArrow* | Number of clicks on down arrow key. |
| *MSecForUpArrow* | Number of milliseconds the user spent on up arrow key. |
| *MSecForDownArrow* | Number of milliseconds the user spent on down arrow key. |

To use variables as features in the models (sections 4.1, 4.2, 4.3), we selected all the variables in the explicit feedback with the exception of 'user_like' because it is the target variable. The purpose of selecting all the variables was to check their impact on predicting user ratings. In implicit feedback however, we selected only 'TimeOnMouse' and 'TimeOnPage', rejecting all others as they were sparse vectors (a vector whose most of the elements are zero).

Table 6.3 : Statistics on explicit feedback

| Feedback | min | Max | Mean | Variance |
|---|---|---|---|---|
| *user_like* | 1 | 5 | 3.35 | 1.68 |
| *Relevant* | 1 | 5 | 3.41 | 1.86 |
| *Novelty* | 1 | 5 | 3.43 | 2 |
| *Readability* | 0 | 1 | 0.87 | 0.11 |
| *Authority* | 0 | 1 | 0.82 | 0.15 |

Table 6.4 : Statistics on implicit feedback

| Feedback | min | max | Mean | Variance |
|---|---|---|---|---|
| *TimeOnMouse* | 0 | 320611 | 1976.4 | 29049887.3 |
| *TimeOnPage* | 0 | 5138378 | 71430.3 | 18683642234 |
| *EventOnScroll* | 0 | 160 | 0.92 | 13.7 |
| *ClickOnWindow* | 0 | 81 | 0.87 | 5.5 |
| *TimeOnHScroll* | 0 | 0 | 0 | 0 |
| *TimeOnVScroll* | 0 | 0 | 0 | 0 |
| *NumOfPageUp* | 0 | 19 | 0.1 | 0.67 |
| *NumOfPageDown* | 0 | 19 | 0.12 | 0.86 |
| *MSecForPageUp* | 0 | 5978 | 17.3 | 32784.86 |
| *MSecForPageDown* | 0 | 7391 | 22.46 | 50489.13 |
| *NumOfUpArrow* | 0 | 23 | 0.08 | 0.52 |
| *NumOfDownArrow* | 0 | 91 | 0.96 | 18.24 |
| *MSecForUpArrow* | 0 | 7378 | 23.26 | 46440.9 |
| *MSecForDownArrow* | 0 | 17743 | 175.1 | 641665.3 |

## 6.3. Data Pre-processing

Data pre-processing is usually the first step of data mining and knowledge discovery. Data pre-processing can have an impact on generalization performance of Machine Learning algorithm [Kotsiantis et al., 2006]. Most of the real world datasets suffer with problems of missing values and ambiguities, similar was the case with our dataset. To deal with these issues in the feedback 'classes', data pre-processing was done. There were a few steps involved in data pre-processing:

- First, we removed all the rows or information on articles which didn't belong to any class. There were 2094 user feedbacks in which article didn't belong to any class at first, after removing all of them we were left with around 8000 user feedbacks and 540 classes.

- Next, we removed ambiguities in the feature 'classes', ambiguities like there were several classes 'USA', 'usa' and 'us' referring to the same country or entity, we replaced all of them with a single class.

- After the second step, we removed all the classes which occurred only once as machine learning requires occurrence of each item at least twice so that one can be used in training and other for testing.

After the third step, we checked again if there was any article still left which didn't belong to at least one class. We stopped when we had removed sufficient amount of ambiguities and each article belonged to at least one class.

As for the other features, we replaced blanks with '0' in variables 'TimeOnMouse' and 'TimeOnPage', for variables like 'readability', 'novelty', 'relevant' and 'authority', blanks and -1 were replaced with lowest value they could have i.e. 1 for 'relevant' and 'novelty', 0 for 'readability' and 'authority'.

After performing all the steps above, we were left with 6625 user feedbacks and 335 different classes.

## 6.4. Feature Extraction

For the reasons explained in section 6.1, we wanted to use conditional probabilities of 'classes' in models like Linear and Logistic Regression, so, we extracted two features from the feature 'classes' namely 'joint conditional probability' and 'mean conditional probability'. The steps involved in the process were:

- First, separate classes into tokens, for example, if an article belongs to classes 'sports' and 'basketball', the two tokens will be 'sports' and 'basketball' (in case if an article belongs to only one class, then there is only one token). This is done for all articles.

- Compute the joint frequency of each token with each 'user_like' i.e. {1,2,3,4,5}, joint frequency is the total number of times a token appears with a particular 'user_like'.

- Compute the frequency for all values of 'user_like'.

- Now, we compute the conditional probability of token given 'user_like' or $P(token|user\_like)$. We do this by dividing joint frequency of token and 'user_like' by the frequency of the particular 'user_like' which has appeared with token. For example,

$$P\ ('sports'|user\_like\ =\ 5)\ =\ \frac{joint\ frequency\ of\ 'sports'\ and\ 'user\_like' = 5}{frequency\ of\ 'user\_like' = 5} \quad (4)$$

- To calculate joint conditional probability, we multiply the conditional probabilities of all the tokens of an article (only if there are multiple tokens, otherwise conditional probability becomes joint conditional probability).

- To calculate mean conditional probability, we take the mean of the conditional probability of all the tokens of an article (only if there are multiple tokens, otherwise conditional probability becomes mean conditional probability).

We also calculated 'TimeOnMouse' and 'TimeOnPage' in seconds to use as features. The new features were named 'TimeOnMouse (sec)' and 'TimeOnPage (sec)'.

## 6.5. Results

To present our results in a concise way, we have assigned our features short aliases. Table 6.5 provides description of each alias.

Table 6.5 : Feature names and their respective aliases

| *Feature name* | *Alias* |
|---|---|
| *Relevant* | RL |
| *Novelty* | N |
| *Readability* | RD |
| *Authority* | A |
| *TimeOnMouse* | TM |
| *TimeOnPage* | TP |
| *joint conditional probability* | JCP |
| *mean conditional probability* | MCP |

Results that we obtained after performing experiments (Linear Regression, Logistic Regression and Naive Bayes) on our pre-processed dataset are shown in the Figures 6.1, 6.2 and 6.3 (these results are for all users, results for individual users are given in Figures 6.5, 6.6 and 6.7).

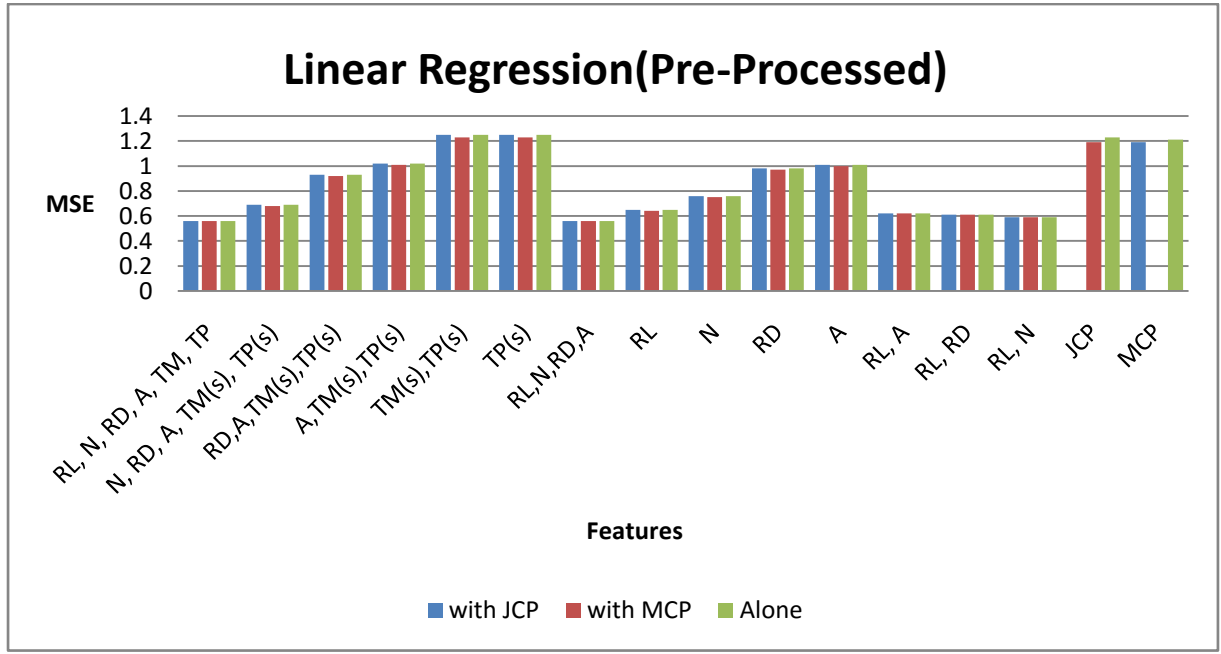

Figure 6.1 : Linear Regression on Pre-Processed Dataset

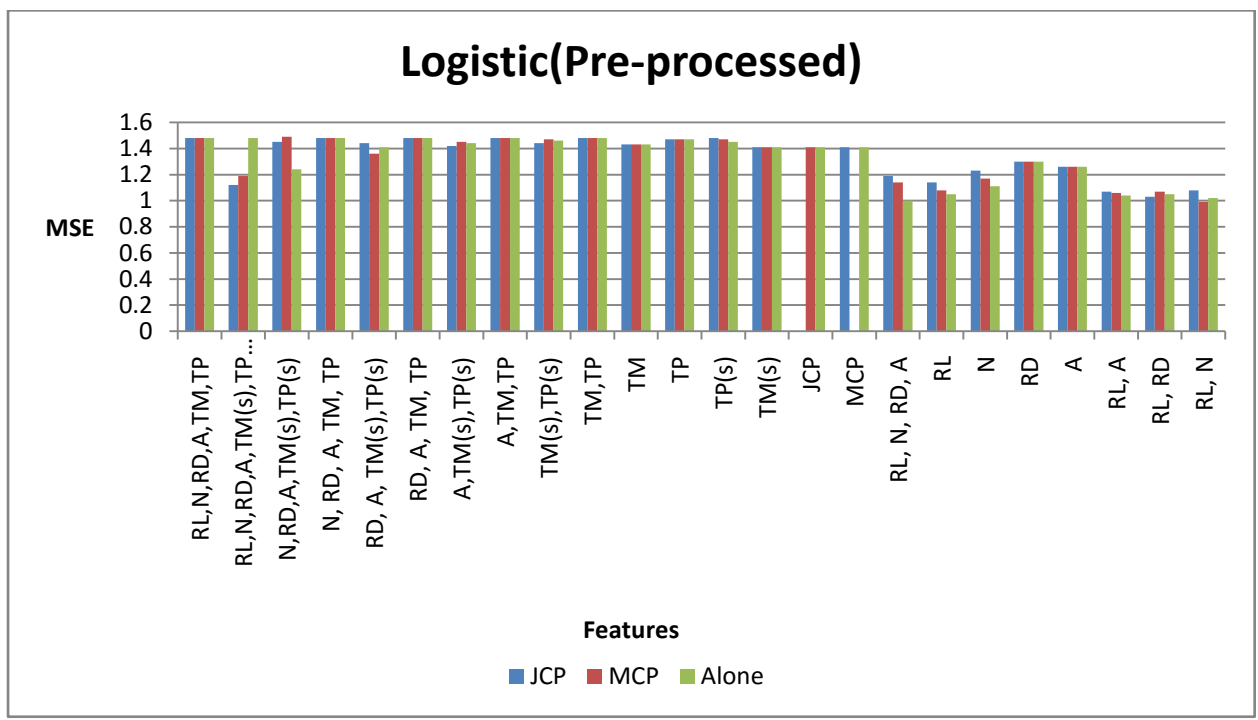

Figure 6.2 : Logistic Regression on Pre-Processed Dataset

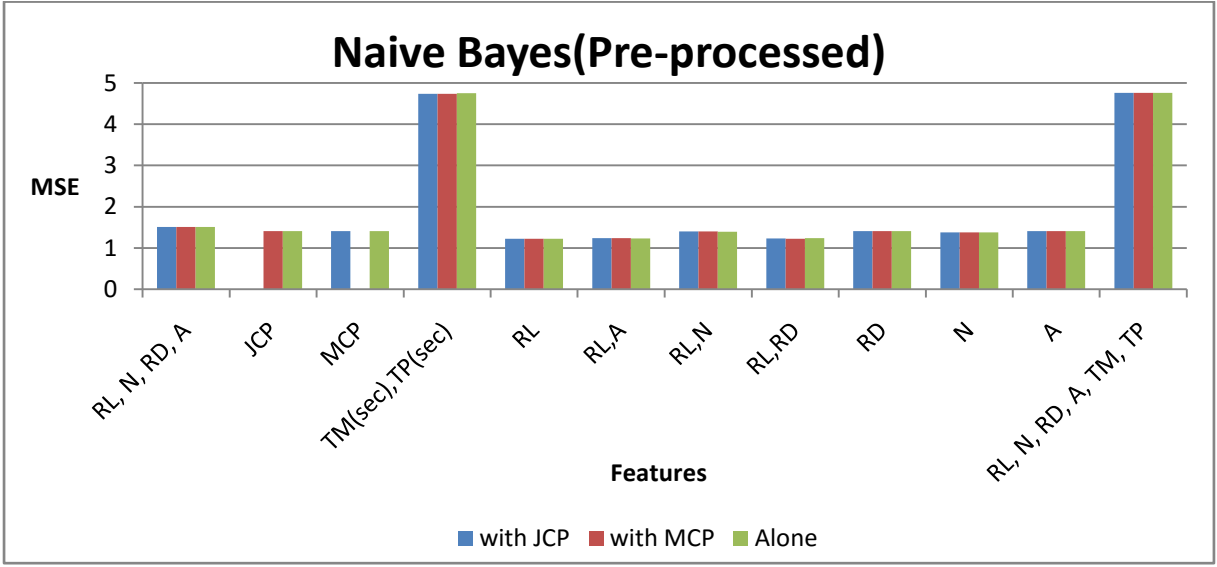

Figure 6.3 : Naive Bayes on Pre-processed Dataset.

After we observed that features 'relevant (RL)' and 'novelty (N)' were performing exceptionally well alone, we decided to compute each feature's correlation with 'user_like', our results are presented in Figure 6.4.

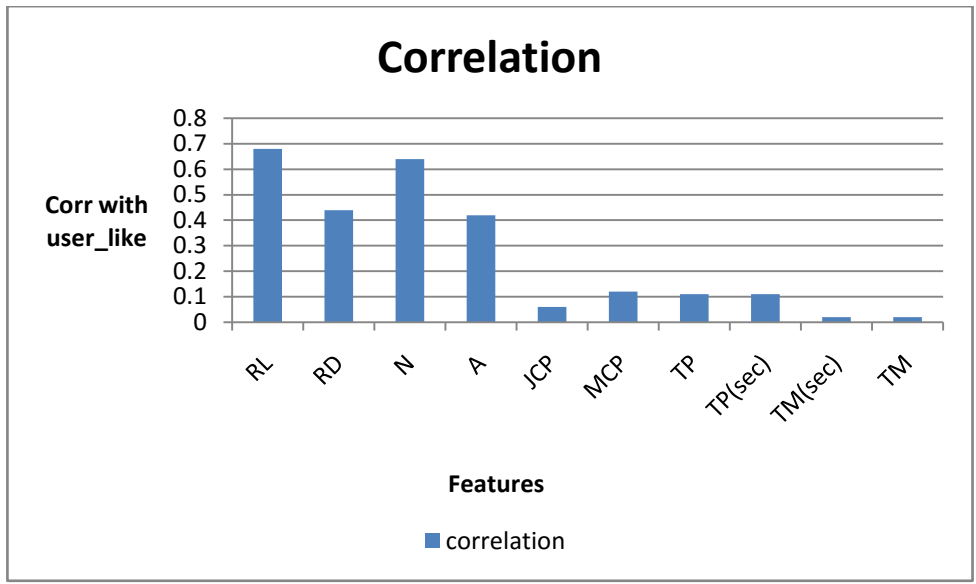

Figure 6.4 : Correlation of user ratings with different features

From Figure 6.4, we can clearly see that features 'relevant (RL)' and 'novelty (N)' are strongly correlated with 'user_like', 'readability (RD)' and 'authority (A)' are somewhat correlated with 'user_like' whereas other features are not at all correlated with 'user_like'. This gives a good indication that features 'relevant (RL)','novelty (N)', 'readability (RD)' and 'authority (A)' will be able to approximate 'user_like' better than other features.

We also have to take into account that all these strongly correlated (with 'user_like') features are explicit feedback, in a real world scenario a user doesn't really give that much explicit feedback and systems have to rely on implicit feedback. As we know that usually a class an article belongs to is given by the system and not the user (in our dataset classes are given by user), so in further experiments we have reduced the number of classes and performed our experiments to see the changes in the performance of features in different models.

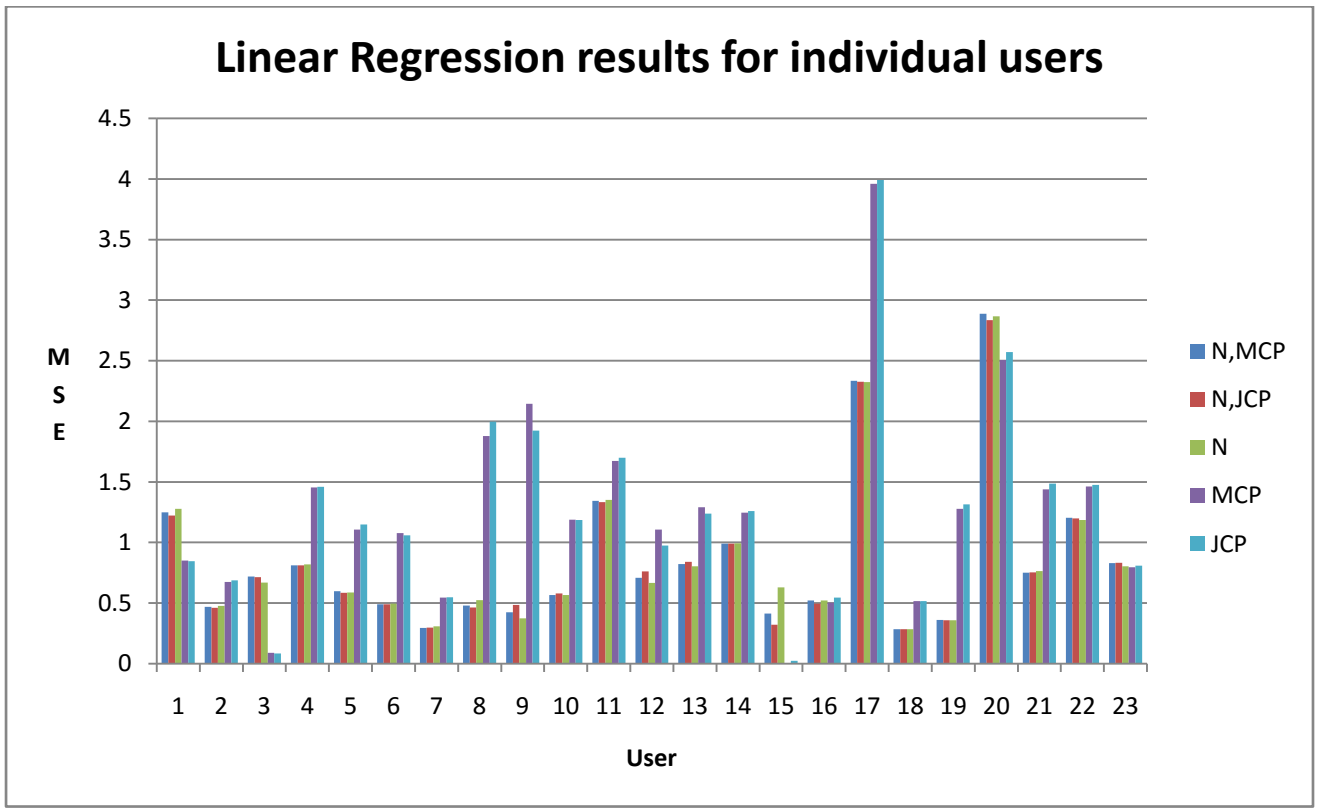

Figure 6.5 : Linear Regression results for individual users

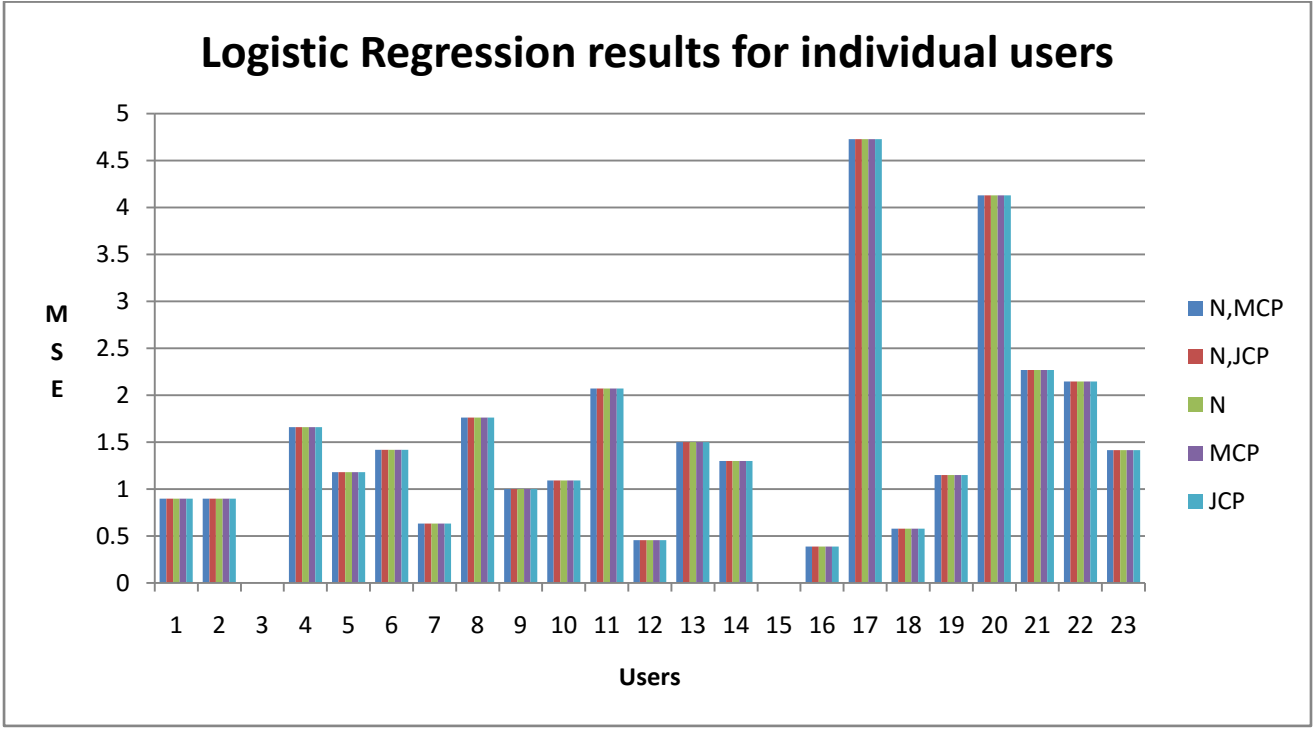

Figure 6.6 : Logistic Regression results for individual users

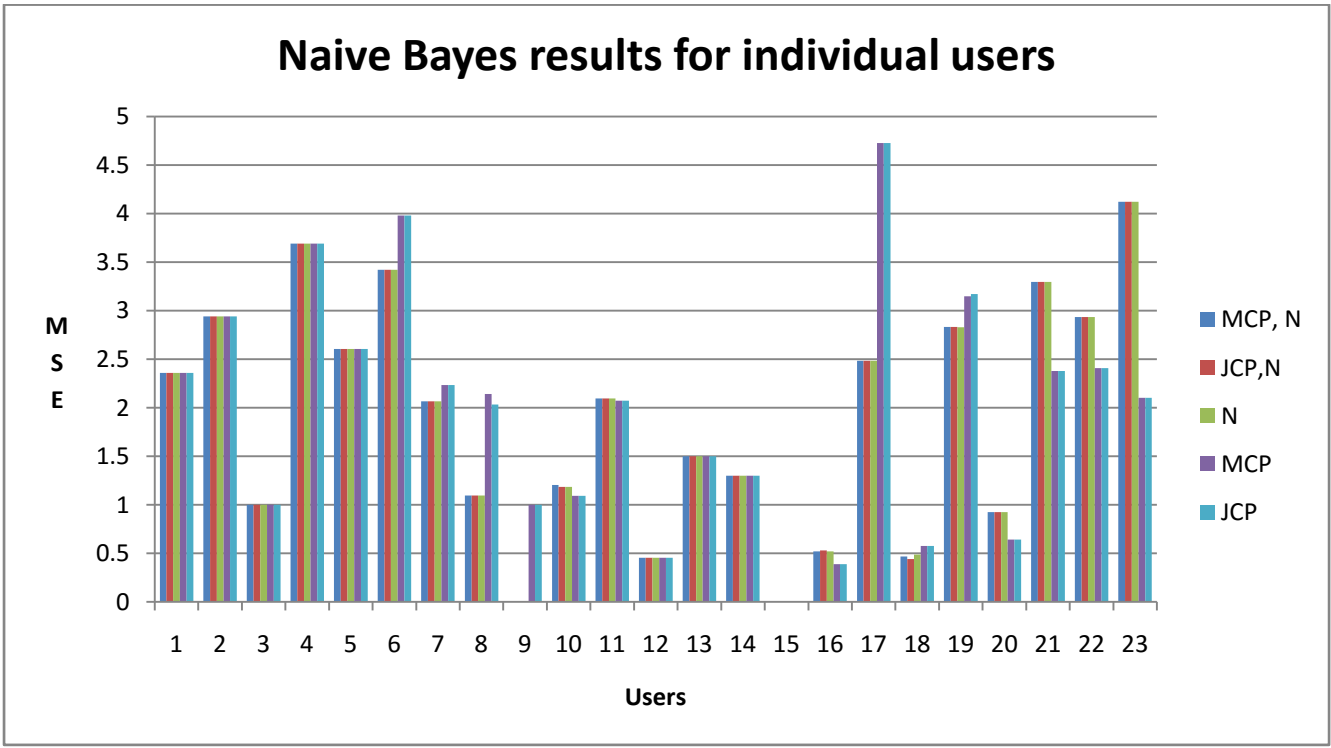

Figure 6.7 : Naive Bayes results for individual users

In Figures 6.1, 6.2 and 6.3 are the results of different models on all users, whereas in Figures 6.5, 6.6 and 6.7 are the results of different models on individual users.

With the different features available to us, models could be tested with many combinations of features, so we decided to go with few combinations in case of all users. But as we know that most of these features are explicit feedbacks and won't be available in a real world data set, so we dropped all of the features and kept only novelty (N) as it is used in recommender systems as evident by [Parra and Amatriain, 2011] and [Parra et al., 2011] . For individual users, we have only shown results using feature novelty (N) and our extracted features i.e. JCP and MCP. For baseline, we used popularity model i.e. recommending the most popular article, this is also a very common practice as evident by [Hu et al., 2008], [Parra and Amatriain, 2011] , [Parra et al., 2011]  and [Johnson, 2014] .

In case of all users, the result was that linear regression model with our extracted features was able to outperform the popularity model or baseline but it was unable to outperform the explicit feedbacks as they were strongly correlated with user_like and our extracted features were not as clear in Figure 6.4.

In case of individual users, linear regression model with our extracted features was able to outperform the popularity model or baseline for most of the users but not all.

We performed experiments by reducing the number of classes too and found out that number of classes was having an impact on performance of our models which contradicts the founding in [Garcin et al., 2012].

Below in Figure 6.8 are the mean square errors we got using only extracted features with different models which shows that linear regression outperforms all other models with our extracted features . And Figure 6.9 shows impact of number of classes (these results are on all users rather than individuals), as the number of classes decrease, performance of models increases (or mean square error decreases).

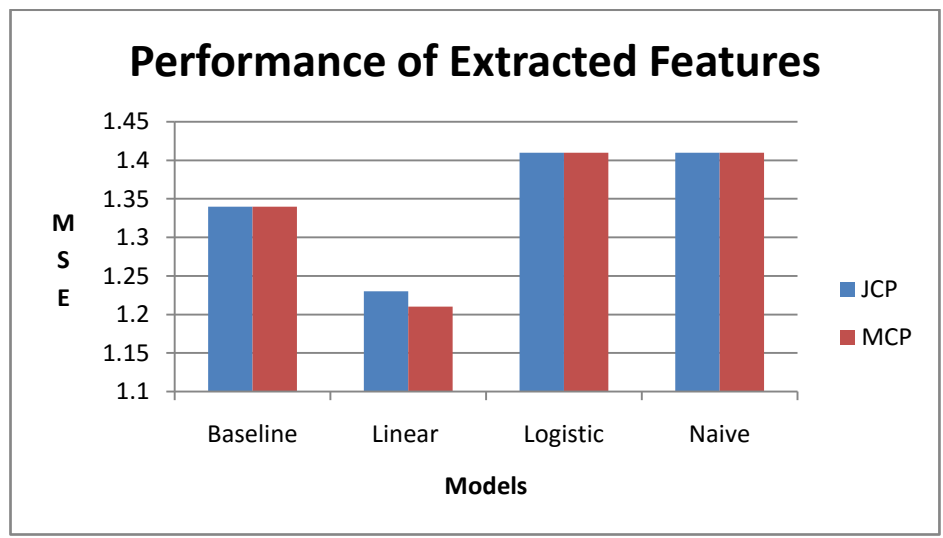

Figure 6.8 : Performance of Extracted Features

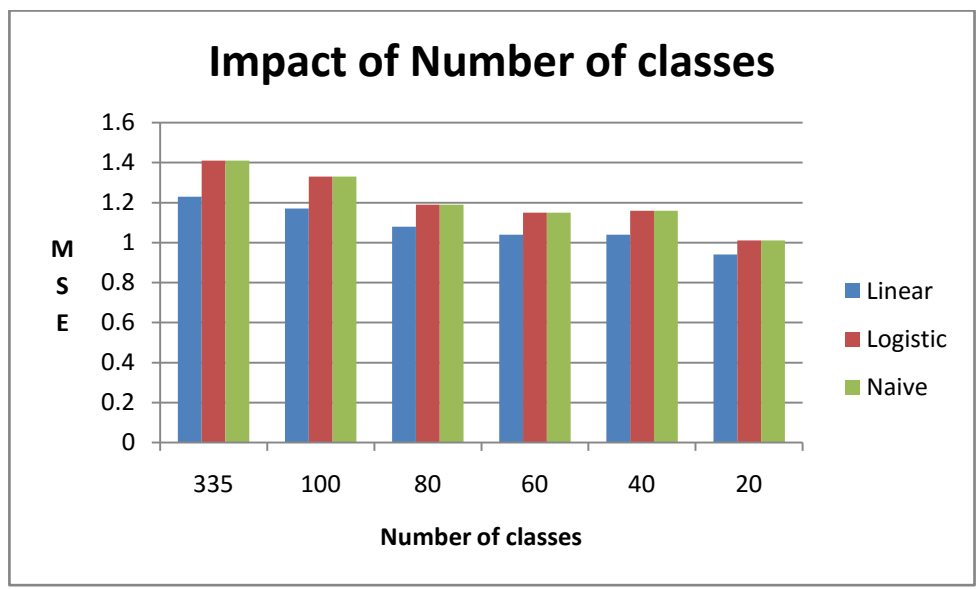

Figure 6.9 : Impact of Number of classes

# 7. Conclusions and Future work

In this thesis, we have shown how the class of an article can be used for news recommendation. We have seen step by step how our system developed. First, we defined and formalized the problem. Then, we pre-processed the data thereby handling ambiguities and missing values. After that, we extracted two features. And lastly, we used the extracted features and features available in the data set to predict how much a user likes a news article.

The most important conclusion that we can draw from our results is that the class of the article can be used to improve a recommendation algorithm. News articles are strongly related to its class or category and this has an impact on users preferences. Using a supervised algorithm we were able to observe the impact of this feature (i.e. articles' class). Out of the three learning models i.e. linear regression, logistic regression and Naive Bayes, linear regression turned out to be the best one.

Extracted features were able to outperform popularity model but they are still not able to outperform explicit feedback like novelty (N), at least not with the three models we have used.

Reducing the number of classes has an impact on performance of every model. As the number of classes decrease, performance of every model increases i.e. number of classes and performance of a model are inversely proportional to each other. This finding is contradictory to finding of [Garcin et al., 2012] where they found no impact of number of classes on performance. The difference could be because in our dataset, 'classes' was an explicit feedback, it is possible that in their dataset the feedback which is similar to 'classes' is implicit feedback.

Extracted features i.e. Joint Conditional Probability (JCP) and Mean Conditional Probability (MCP) though performed reasonably well alone, average of mean square error on all datasets was around 0.98 for linear regression and around 1.14 for logistic regression and Naive Bayes, however they were unable to have any influence on other features.

Naive Bayes turned out to be the weakest among three models as it performed worse with TimeOnMouse (TM) and TimeOnPage (TP), with average of mean square error being above 4. The bad performance of Naive Bayes was not surprising because Naive Bayes is not a robust classifier, and the problem in hand requires a model that is able to detect correlations that are not easily detected with small datasets

In future work, we would like to test our approach on a dataset where classes are implicit feedback rather than explicit feedback thereby performance would be more real world and we would also be able to observe any difference in performance of implicit versus explicit feedback.

There are many other options of algorithms that perform supervised learning. In the present thesis, we have worked with only three of them, however it would be interesting to evaluate others. Another interesting approach would be to solve the problem of news recommendation as unsupervised learning problem. For example, models that are able to detect underlying variables that are enclosed in the news articles (i.e. LDA).

However, our approach is one of the many approaches and in the future we would like to propose different and more complex approaches with ultimate goal of designing a news recommendation system which are able to improve the state-of-the-art.